# How do software citation formats evolve over time? A longitudinal analysis of R programming language packages

Yuzhuo Wang (Anhui University, China)

Kai Li (University of Tennessee, Knoxville, USA) kli16@utk.edu

**Abstract:**

Under the data-driven research paradigm, research software has come to play crucial roles in nearly every stage of scientific inquiry. Scholars are advocating for the formal citation of software in academic publications, treating it on par with traditional research outputs. However, software is hardly consistently cited: one software entity can be cited as different objects, and the citations can change over time. These issues, however, are largely overlooked in existing empirical research on software citation. To fill the above gaps, the present study compares and analyzes a longitudinal dataset of citation formats of all R packages collected in 2021 and 2022, in order to understand the citation formats of R-language packages, important members in the open-source software family, and how the citations evolve over time. In particular, we investigate the different document types underlying the citations and what metadata elements in the citation formats changed over time. Furthermore, we offer an in-depth analysis of the disciplinarity of journal articles cited as software (*software papers*). By undertaking this research, we aim to contribute to a better understanding of the complexities associated with software citation, shedding light on future software citation policies and infrastructure.

# 1. Introduction

Software has become a cornerstone of the information society and has been broadly adopted in all societal sectors (United Nations Conference on Trade and Development, 2012; Manovich, 2013; Branstetter et al., 2019). This is particularly so in the research communities: under the data-driven research paradigm, software plays central roles in almost every step of scientific activities, including data collection, processing, analysis, and visualization (Edwards et al., 2013; Schindler et al., 2020; Borgman et al., 2012; Howison et al., 2015). Accordingly, software has gained the status of a "first-class research object" (Chassanoff & Altman, 2020), i.e., a type of object that is as important as published journal articles and requires more visibility and care-taking in the research system.

With the rise of the open science movement and software in the research system, a number of efforts have been made to improve the infrastructure that renders research software more visible and reusable. One notable example is the revision of the FAIR Principles (Wilkinson et al., 2016) to support research software, i.e., FAIR4RS (Lamprecht et al., 2020), which includes 17 principles under the four categories of *findable*, *accessible*, *interoperable*, and *reusable*. These principles highlight the importance of using standardized descriptive metadata to describe software entities in research outputs, a theme also central to other similar efforts, such as the Software Citation Principles developed by the Force11 Software Citation Working Group (Smith et al., 2016). In addition to these higher-level principles, researchers have proposed more specific software citation recommendations, often at the level of the metadata field (Hong et al., 2019a, 2019b; Katz et al., 2021). These principles provide guidelines for various stakeholders to supply and harvest information about software used in academic research (Bouquin et al., 2020; Du et al., 2022).

Many of the principles above are directly related to publishing research software. Borrowing from the concept of *data publication* (Kratz & Strasser, 2014; Parsons & Fox, 2013), *research software publication* can be defined as the process of making the software available, giving it adequate documentation, and rendering it officially citable. As a defining characteristic of the concept, officially citable is one of the most basic prerequisites for software to be integrated into the research system and reused by other researchers (Katz et al., 2021). Accessible citation formats ensure the visibility of software in research outputs, promoting transparency and replicability of research (Howison & Bullard, 2015). However, as a highly dynamic and traditionally marginal object in the research space, software is cited very differently from publications. For example, empirical evidence has shown that software is inconsistently cited: informal mentions of software are common in academic papers (Howison & Bullard, 2015; Pan et al., 2016, 2018), and the multiple levels of software entities (especially in the cases of programming languages) as well as their different versions can lead to inconsistent decisions by researchers about what and how to cite (Li et al., 2017). More profoundly, there may be multiple citation formats that can represent the same software (Li, Chen, & Yan, 2019). All these issues make software citation a formidable problem to solve before a new infrastructure can be developed for data-driven research.

The last issue mentioned above, i.e., multiple citation formats for the same software entity, is especially evident in recent years because of the rising genre of software paper, i.e., an academic, peer-reviewed document to describe a software entity (Chue Hong et al., 2013). Debates exist around the usefulness of software papers as a citable representation of software entities. On the one hand, such papers are regarded as a useful means of improving the findability and reusability of research software (Chue Hong et al., 2013) and to comply with other high-level software citation principles. On the other hand, the software paper is a monolithic and static representation of software entities, which contrasts with the dynamic nature of software (Jay et al., 2021). Beyond this debate, it is evident that software papers and software repositories can compete for the space of software citation, leading to inconsistencies in the representation of research software in scientific publications (Li, Chen, & Yan, 2019).

Given the diverse ways in which software is cited in publications, it is important to establish an overview of the software citation landscape based on empirical data, as a supplement to the higher-level software citation principles. One barrier to this research agenda is that many software citation formats evolve over time. As indicated by our previous work (Li et al., 2017), the diversity of software citation formats is very hard to capture in existing empirical studies, which are often based on either individual software entities or identifying software name mentions in publications.

To bridge the above gap, this research aims to analyze how the citation formats of all R-language packages changed between August 2021 and October 2022. We examined all software packages available in the CRAN (Comprehensive R Archive Network)[1] software repository. Despite the existence of other R-language software repositories such as Bioconductor (Gentleman et al., 2004), CRAN serves as the official and the largest software package repository of the R-language ecosystem (Hornik, 2012). For this reason, we regard this sample of software entities as a comprehensive representation of all R-language packages. More specifically, we conducted a longitudinal analysis using the official citation formats collected from CRAN in 2021 and 2022 to understand the following research questions:

**RQ1: What document types are used in the software citation format?** This question aims to understand the types of objects representing software entities in the citation format, especially in terms of the dichotomy between software papers and software projects reported in earlier research (Li, Chen, & Yan, 2019). This variation in document types and its connection to the nature of research software have rarely been examined in the literature. We classified the objects underlying the citation format using the object type tag and textual information offered by the software developers and further explored how such information changed between the two time points. This question sheds fresh light on the diversity of software citation practices in terms of the underlying cited object.

**RQ2: How do metadata elements in software citation formats evolve over time?** Beyond the document types of citation formats, we also aim to compare the metadata elements in citation formats of the same software object to understand whether there are other changes in the citation format over time. This question will help us better understand the fluidity of software citation formats, which has not been fully documented in the literature, using large-scale empirical data.

**RQ3: In what venues are software papers published?** Based on RQ1, we further examine software papers cited in our sample, focusing on the journals and research disciplines of the publications. By doing so, we hope to conduct a first scanning of the landscape of software papers, given the lack of prior large-scale empirical analyses on this topic. The results from this question will also help future studies to identify software papers from the bibliographic universe and better trace activities of research software development and reuse in academic communities.

This article is the first study to examine the evolution of software citation formats using a large-scale data sample. Our research provides a new perspective from which to understand the diversity of software citation formats in the R programming language and how the landscape shifts over time. This knowledge contributes to a deep understanding of the relationship between research software and scientific publications and has strong policy and methodological implications.

---

[1] https://cran.r-project.org/index.html

## 2. Related work

Software and its computational capacities have been commonly regarded as a central driver of knowledge production in contemporary sciences (Kelty, 2001; Wolfram, 1984). Under the data-driven paradigm of scientific research, software is involved in almost every step of the research workflow (Edwards et al., 2013). The increasing importance of software to the research system has inspired recommendations from the research community to how to publish and cite research software, so that research can be more effectively reproduced, the software entities can be reused by other researchers, and software developers can get credit for their work (Chassanoff & Altman, 2020).

Against this backdrop, the concept of *research software publication* has been proposed to describe the process of making the software available, giving it adequate documentation, and rendering it officially citable, practices with clear parallels to the concept of research data publication (Kratz & Strasser, 2014; Parsons & Fox, 2013). Despite the significant differences between research data and software in the context of scientific citation (Katz et al., 2016), it is acknowledged that both types of objects should be covered by the FAIR Principles, hence the FAIR Principles for Research Software, or FAIR4RS (Barker et al., 2022; Lamprecht et al., 2020). This revised version of the FAIR Principles stresses the roles of metadata descriptions of software in making these objects transparent and reusable in scientific research, which highlights the importance of citation and documentation to integrating software into the research system.

Despite higher-level guidance such as FAIR4RS and the Software Citation Principles (Smith et al., 2016), empirical evidence is still needed to better understand the landscape of software citation and reuse, as the first step towards a future research infrastructure that can fully support research software. Existing empirical research has identified various limitations in how research software is represented in scientific publications, such as the frequent informal mentions of software in academic papers (Howison & Bullard, 2015; Pan et al., 2016, 2018), the hierarchical structure of programming language systems that is difficult to represent as citations and may result in citation inconsistencies (Li et al., 2017) and multiple citation formats that may be used for the same software (Li, Chen, & Yan, 2019). All these issues have posed significant barriers to empirical studies on research software using large-scale citation data.

More recently, researchers have moved to using automatic methods to extract software name mentions (Li & Yan, 2018; Wang & Zhang, 2020) in an effort to circumvent the limitations mentioned above. Specifically, they have published large-scale software mention datasets (Du et al., 2021; Istrate et al., 2022) and used contextual information of such mentions to construct knowledge graphs (Druskat, 2020; Kelley & Garijo, 2021; Schindler et al., 2022), as a part of the growing trend to use complex network and data science technologies in quantitative studies of science (Manghi et al., 2021).

Despite this progress in empirical studies on research software, software citations in reference list still represent a much more abundant—and hence more accessible and comprehensive—data source than software mention in full-text publications for understanding the impacts of research software. Consequently, understanding the practice of software citation remains a critical research question for the next step in this line of research, which is a major motivation of the present study.

## 3. Method

### 3.1 Data collection

In this study, we analyzed the citation formats of R packages included in the official software repositories of the R programming language, the Comprehensive R Archive Network (CRAN). Created in 1996, CRAN is composed of a network of FTP (File Transfer Protocol) and web servers that store current and previous versions of code and documentation for R software and its packages. The R core team, made up of teams from the Institute for Statistics and Mathematics at the Vienna University of Economics and Business, is responsible for the operation of CRAN[2]. Specifically, the team processes submissions to CRAN, provides CRAN package checks on combinations of various platforms, and maintains the package repository.

We downloaded the citation format of all R packages based on the full package list[3] available from CRAN in August 2021 and October 2022, respectively. Using the two lists, we installed all packages and used the *citation()* function in R to retrieve the official citation formats assigned by the developers (LaZerte, 2021). The citation format for R packages is presented as a BibTeX entry, which includes all metadata elements that compose the citation format. Table 1 shows three examples of the citation format corresponding to the three types of citable objects (see section 3.2).

**Table 1. Examples of different kinds of citation formats**

| Type | *Manual* | *Article* | *Other* |
|---|---|---|---|
| **Package** | intensitynet | plantecophys | harmonicmeanp |
| **Title** | IntensityNet: analysis of spatial point patterns occurring in networks | Plantecophys – an r package for analyzing and modeling leaf gas exchange data | The harmonic mean p-value for combining dependent tests |
| **Author** | Pol Llagostera Blasco, Matthias Eckardt | Remko A. Duursma | Daniel J. Wilson |
| **Journal** | — | PLoS ONE | Proceedings of the National Academy of Sciences U.S.A. |
| **Publisher** | — | — | National Academy of Sciences U.S.A. |
| **Year** | 2021 | 2015 | 2019 |
| **URL** | — | — | https://www.pnas.org/content/116/4/1195 |
| **DOI** | — | 10.1371/journal.pone.0143346 | |
| **ISBN** | — | — | |
| **Version** | 1.3.1 | 1.4.6 | 3 |

We also found cases where one BibTeX entry contained multiple citations of the same package. This situation accounts for 18.3% of all packages in our 2022 sample. The existence of multiple citations makes it difficult to compare between the two samples, so we kept only the first citation format in these files, which is supposedly the default citation format researchers use.

---

[2] https://www.r-project.org/foundation/board.html
[3] https://cran.r-project.org/web/packages/available_packages_by_name.html

The description of these two package samples is shown in Table 2. There are packages in each sample with no citation information; these were removed from subsequent steps of the analysis given the small number of such packages.

**Table 2. Description of the two samples of R packages**

| Sample | Packages | Packages with citation format |
|---|---|---|
| 2021 | 17,985 | 17,377 |
| 2022 | 18,700 | 18,160 |

We classified all these packages into three groups based on their presence in the two samples:

- *Removed*: packages found only in the 2021 sample;
- *Added*: those found only in the 2022 sample;
- *Remained*: those in both samples.

The composition of our analytical sample by year and category is shown in Table 3.

**Table 3. Composition of package categories by year**

| Category | 2021 sample | 2022 sample |
|---|---|---|
| *Removed* | 1,759 | 0 |
| *Added* | 0 | 2,542 |
| *Remained* | 15,618 | 15,618 |
| **All** | 17,377 | 18,160 |

### 3.2 Classification of cited objects

Using the document type tag offered by the software developers, we reclassified all document types available from downloaded data into the following three groups, whose mapping and descriptive analysis are shown in Table 4:

- *Article*: published journal articles
- *Manual*: software repository pages
- *Other*: all other categories (see breakout in table)

**Table 4. Mapping of package categories by year**

| New category | Original category | Count (2021) | Count (2022) |
|---|---|---|---|
| *Article* | Article | 2,102 | 2,260 |
| *Manual* | Manual | 14,805 | 15,450 |
| *Other* | Preprint | 183 | 156 |
| | Misc | 95 | 107 |
| | Book | 58 | 67 |
| | InProceeding | 43 | 30 |
| | Unpublished | 33 | 42 |
| | TechReport | 27 | 20 |
| | InCollection | 7 | 7 |
| | Proceedings | 7 | 4 |
| | InBook | 6 | 6 |
| | PHDThesis | 6 | 7 |
| | MasterThesis | 5 | 4 |

The selection of the top two categories is informed by the dichotomy between software papers (*Article*) and software repositories (*Manual*), which are the most frequently cited objects in software citations (Li, Chen, & Yan, 2019). A few points should be clarified regarding this classification. First, some unpublished academic documents are assigned the *Article* type by software developers. Given the fact that they are not published, we identified them by matching phrases such as "in press," "accepted," "under review," or the name of preprint servers, such as "ArXiv," "BioXiv," and "preprint" from the publication information in the citation format. These identified preprint publications are renamed *Preprint* and classified as *Other*. Second, we are only focusing on journal publications, rather than other types of academic documents in the list (mostly in the *Other* category), because (1) journal articles are easier to classify into knowledge domains than conference publications and (2) the size of this category is much larger than others, such as *Book* and *InProceedings*. We fully acknowledge that software publications can also take the forms of books and conference papers. However, given the small sizes of these categories, it is not very impactful to include them in this analysis.

### 3.3 Disciplines of software papers

We regard all citations in the *Article* category in our analytical sample as software papers. For these, we first extracted the journal names, which were used to acquire the WoS categories representing the discipline of the publications. To better understand the distribution of software papers across research domains on a higher level, we further mapped the WoS categories to the six knowledge domains defined in the GIPP Research Areas schema[4], including *Arts & Humanities*; *Clinical, Pre-Clinical & Health*; *Engineering & Technology*; *Life Sciences*; *Physical Sciences*; and *Social Sciences*. The mapping was achieved using the official mapping table supplied by Clarivate.[5] A total of 1,771 and 1,804 papers in the respective subsamples were mapped to the scheme, accounting for more than 80% of all software papers.

Journals classified into multiple GIPP knowledge domains after mapping were assigned to research disciplines using fractional counting. Despite the limitations of using journal-level classification to represent the discipline of individual publications (Shu et al., 2019), we believe this method can offer firsthand knowledge of how software papers are distributed across knowledge domains.

### 4. Results

### 4.1 Types of R software package citation format

This section focuses on how the three types of objects—*Article*, *Manual*, and *Other*—are used in citation formats. As shown in Table 5, only about 12% of all R packages use a published article as the citation format. Arguably, these packages are more likely than their counterparts to be research software. Moreover, there is a small rise in the use of *Article* citations and the opposite trend for *Manual* over time, but such changes are very slight during this 15-month window.

---

[4] http://help.prod-incites.com/inCites2Live/filterValuesGroup/researchAreaSchema/gippDetail.html
[5] http://help.prod-incites.com/inCites2Live/indicatorsGroup/aboutHandbook/appendix/mappingTable.html

Table 5. Composition of citation format categories by year

| Category | 2021 ($n$ = 17,377) | | 2022 ($n$ = 18,160) | |
|---|---|---|---|---|
| | Count | Percent | Count | Percent |
| *Article* | 2,102 | 12.10% | 2,260 | 12.44% |
| *Manual* | 14,805 | 85.20% | 15,450 | 85.08% |
| *Other* | 470 | 2.70% | 450 | 2.48% |

To further contextualize the results in Table 5, we compared four subsets of the two samples, i.e., the *Removed*, *Remained* (in each year), and *Added* categories for each sample. As shown in Table 6, *Added* packages have the lowest share of *Article* citations but the highest share of *Manual* citations. An apparent reason behind this difference is that journal articles take time to be published. Hence, newer packages are more likely to use the data repository as the citation format. On the same page, in the *Removed* category, the share of *Article* citations is also lower than in the two *Remained* groups, which indicates that packages removed from the software repository tend *not* to be those with more investment, i.e., those with a published article. Between the two *Remained* groups, from 2021 to 2022, the share of *Article* formats increased, albeit very slightly. This also shows a trend in which more software developers choose to publish software papers for their software packages, at least those with academic value, which partly demonstrates the increasing embrace of the emerging genre of software papers. It should be noted that 36 of the new software papers in 2022 were already deposited on a preprint server in 2021.

Table 6. Composition of citation format categories by packages

| Category | Removed | | Remained (2021) | | Remained (2022) | | Added | |
|---|---|---|---|---|---|---|---|---|
| | Count | Percent | Count | Percent | Count | Percent | Count | Percent |
| *Article* | 132 | 7.50% | 1970 | 12.61% | 2118 | 13.56% | 142 | 5.59% |
| *Manual* | 1,595 | 90.68% | 13,210 | 84.58% | 13,118 | 83.99% | 2,332 | 91.74% |
| *Other* | 32 | 1.82% | 438 | 2.80% | 382 | 2.45% | 68 | 2.67% |
| All | 1,759 | 100% | 15,618 | 100 % | 15,618 | 100% | 2,542 | 100% |

The different compositions of the *Remained* packages in Table 6 show that the document type of the same software package can change over time. Among all packages in the *Remained* group, 403 experienced a change in citation format type, revealing that the identity of software objects being cited can change dramatically over time, even though the share of such changed citations is relatively small.

We analyzed these changes in further detail to understand them better. Figure 1 shows how the document types changed for all 403 packages, including 27 papers with an identical citation title between the two samples. A notable trend in the graph is that the majority (57.07%) of these changes are from an unpublished source (either *Manual* or *Other*) to a published software article (*Article*). These, together with 54 citations moving from *Manual* to *Other*, indicate a strong trend toward the publication of software papers over time.

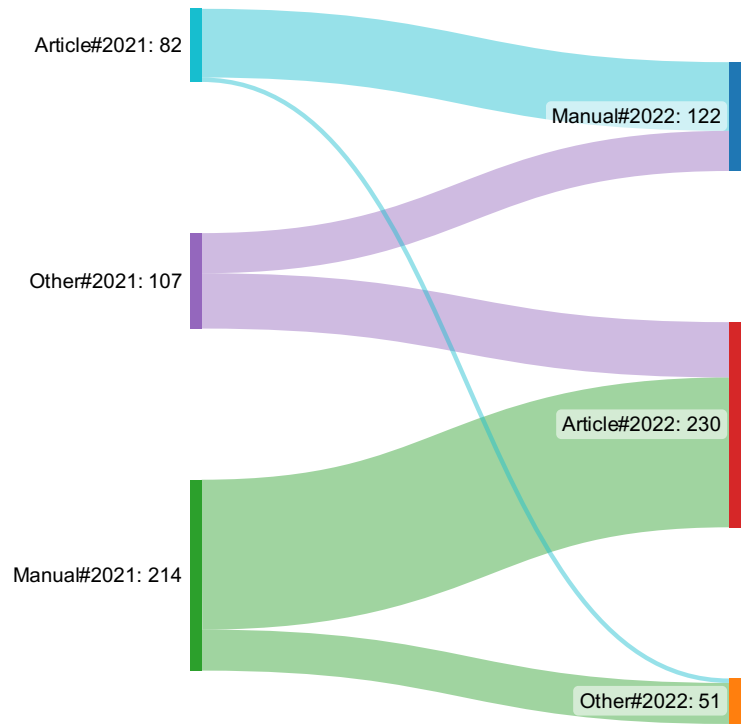

**Figure 1. Change of document types from 2021 (left) to 2022 (right)**

However, movement was also observed in the other direction. Notably, 82 packages moved from *Article* to the other two categories in our dataset; most of these use a repository page in 2022. Among all these packages, only a small fraction (19 of 82) are connected to the existence of multiple citation formats, where a repository format was assigned as the first format in 2022, despite the fact that a data paper was used before. For the rest, this change occurred only because a repository replaced the software paper used previously. This shows the complexity of the changes that software packages' citation formats may undergo; future studies are warranted to fully understand the reasons behind such changes.

**4.2 How are other metadata elements in the citation format change over time?**

This section investigates how the metadata elements in the citation formats change over time. 9,423 *Remained* packages (or 60.33% of all *Remained* packages) have an updated format in 2022 compared to 2021, which shows that software citation format choices are quite fickle, corresponding to the dynamic nature of software entities and posing a major challenge to software citation.

We focus specifically on the 9,020 *Remained* packages that exhibit changes in citation format while retaining the same document type between the two samples, as those with type changes have been analyzed in section 4.1. The distribution of these packages across the three types of cited documents is similar to that of the larger sample (as shown in Table 5), with 11.31% in *Article*, 86.70% in *Manual,* and 1.99% in *Other*.

Table 7 presents the changes in metadata elements for the *Article* and *Other* categories; in both cases, changes are predominantly seen in the software version. This is consistent with the consensus that version is an important element for identifying software, yet difficult to integrate into existing citation practices and infrastructure (Smith et al., 2016). However, the rapid changes in software versions in our results are mitigated by the fact that in practice, the version of R packages is often neglected by researchers providing citations (Li et al., 2017).

**Table 7. Share of *Article* and *Other* citations with changes in metadata elements**

| Type | *Article* | | *Other* | |
| --- | --- | --- | --- | --- |
| | Count | Percent | Count | Percent |
| Version | 1014 | 99.41% | 180 | 100.00% |
| URL | 159 | 15.59% | 17 | 9.44% |
| DOI | 96 | 9.41% | 5 | 2.78% |
| Year | 28 | 2.75% | 10 | 5.56% |
| Title | 25 | 2.45% | 4 | 2.22% |
| Other | 50 | 4.90% | 8 | 4.44% |

Compared to the first two citation categories, more metadata elements changed in *Manual* citations, as shown in Figure 2. The top three metadata elements that changed are the URL, version, and publication year of the package, each of which appears in more than 50% of all citations in the *Manual* category. As in the case of version discussed above, not every metadata element in the list is equally important for the final citation format. For example, the author number is hardly used in the citation string, despite the fact that it is included in the BibTeX citation file. However, these top three categories are important components in the citation format of R packages, pointing to key identifying information related to the software entities. As a result, the large share of citations that changed during the one-year window represent a distinct situation in software citation that should be addressed in future research infrastructure.

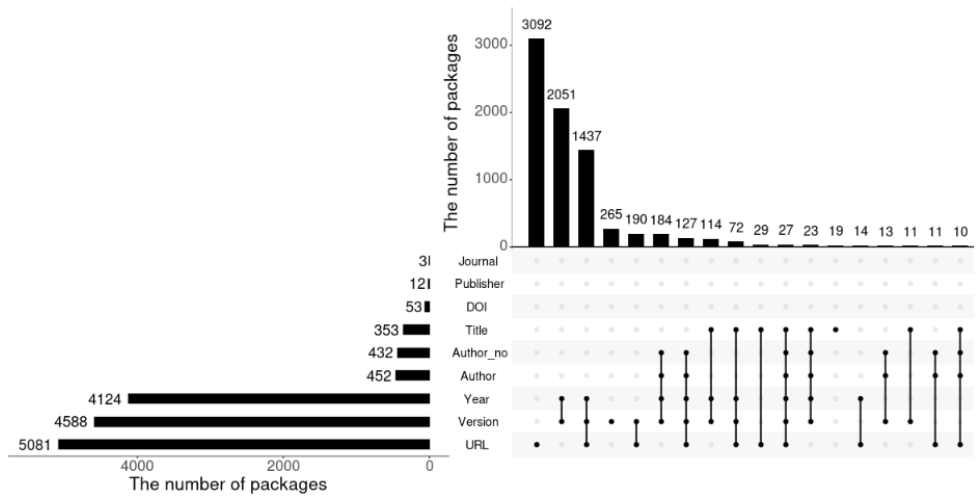

**Figure 2. Share of updated elements in *Manual* citations with unchanged document type**

Of the 9,020 packages examined above, 382 (4.24%) have different titles across their citations in the two samples. In current citation practice, the title is the most important identifying information for the cited object; hence, we regard this as the most significant change. We examined the reasons behind such changes and found that in the *Article* category, all 25 instances under this scenario were due to the introduction of a new software publication to replace the old publication as the citation object, similar to the case of *Other*. However, *Manual* citations can have a new title without the introduction of a new resource. For example, the citation title of the package *enrichwith* in 2021 was "enrichwith: Methods to enrich list-like R objects with extra components," and the title was updated to "enrichwith: Methods to Enrich R Objects with Extra Components" in 2022, but the other elements of the citation format remained identical. However, in all cases, a change in the citation title presents a major barrier to the tracing of software entities using citation data, as demonstrated by our previous study (Li, Chen, & Yan, 2019).

We further calculated the age of software citation formats by subtracting the publication year of the citation from the time of data collection, with results shown in Table 8. The *Manual* category alone contributes to the decreasing trend from 2021 to 2022, whereas the other two categories exhibit greater citation age after a year. The overall decrease in citation age shows the volatile nature of the citation age of software entities, especially in the case where software is represented as a software repository. This is because updates to the citation format can change how old the software is, and we cannot assume a stable connection between a software object and its citation year, as we can in the case of scientific publications.

**Table 8. Average age of citation formats by category**

| Category | Removed | Remained (2021) | Remained (2022) | Added | All_2021 | All_2022 |
|---|---|---|---|---|---|---|
| *Article* | 5.81 | 5.55 | 6.03 | 3.05 | 5.56 | 5.84 |
| *Manual* | 5.04 | 1.93 | 2.16 | 0.20 | 2.26 | 1.87 |
| *Other* | 4.55 | 4.13 | 5.03 | 1.38 | 4.15 | 4.47 |
| **All** | 5.09 | 2.44 | 2.75 | 0.39 | 2.71 | 2.42 |

Among the citations corresponding to the four different types of packages, those for the *Added* packages are significantly younger, while *Removed* package citations are the oldest, which shows a clear succession of the old by the new. However, against this overall trend, one *Added* package, *QHScrnomo*, introduced the paper "A competing-risks nomogram for sarcoma-specific death following local recurrence," published in 2003, in its 2022 updates. This example shows a complex relationship between software packages and publications that cannot be taken for granted.

**4.3 Disciplines of R software papers**

We further analyzed the disciplinary classification of journals where R software papers were published. Table 9 presents the top 10 journals appearing in our two subsamples, which are a mixture of journals dedicated to software papers (such as the *Journal of Statistical Software* and *The R Journal*) and domain-focused journals (such as *Bioinformatics* and *PLoS ONE*). The *Journal of Statistical Software* has a particularly strong connection with R packages in the sense that the majority of its publications are dedicated to R packages. This is partly because of the special position of R among all open-source statistical software, though Python has a much smaller presence in the journal (Fox & Leanage, 2016). Beyond the *Journal of Statistical Software*, there are venues dedicated to publishing R packages, such as *The R Journal* and *R News*, which themselves are not indexed in major research databases. However, as demonstrated by the table, these more specialized journals also play fundamental roles in connecting R packages to the research system. Another important observation from the table is that the landscape of R software journals largely remains consistent from 2021 to 2022: the top journals remain the same, with only a few differences in ranking.

**Table 9. Top 10 R software journals from two subsamples**

| 2021 sample | 2022 sample |
|---|---|

| Journal | Count | Percent | Journal | Count | Percent |
| --- | --- | --- | --- | --- | --- |
| Journal of Statistical Software | 653 | 30.95% | Journal of Statistical Software | 684 | 29.65% |
| The R Journal | 122 | 5.78% | The Journal of Open Source Software | 134 | 5.81% |
| The Journal of Open Source Software | 112 | 5.31% | The R Journal | 122 | 5.29% |
| Bioinformatics | 107 | 5.07% | Bioinformatics | 113 | 4.90% |
| Methods in Ecology and Evolution | 86 | 4.08% | Methods in Ecology and Evolution | 103 | 4.46% |
| PLoS ONE | 32 | 1.52% | PLoS ONE | 31 | 1.34% |
| BMC Bioinformatics | 30 | 1.42% | BMC Bioinformatics | 31 | 1.34% |
| R News | 30 | 1.42% | R News | 30 | 1.30% |
| Computational Statistics & Data Analysis | 22 | 1.04% | Computational Statistics & Data Analysis | 28 | 1.21% |
| Ecography | 20 | 0.95% | Ecography | 20 | 0.87% |

We further examined the knowledge disciplines of the journals in our sample based on their WoS classifications as mapped to the GIPP scheme. We used fractional counting for journals and publications classified into multiple knowledge domains, as seen in Table 10. In general, most of the R software papers were published in STEM fields. Life Sciences, Physical Sciences, and Engineering & Technology account for over 80% of journals and 90% of all software papers in the sample. However, the fact that Life Sciences and Physical Sciences (instead of Engineering & Technology) are the top domains for R publications suggests that R packages and papers are often contributed by researchers from application knowledge domains, instead of computer science and other engineering fields. On the other end of the spectrum, there is just one paper, entitled "How to capitalize on a priori contrasts in linear (mixed) models: A tutorial," published in an *Arts & Humanities* journal. This paper was published in the *Journal of Memory and Language* for the package *HYPR*, which illustrates the diversity in where research software is developed.

**Table 10. Fractional number of journals and papers in six domains**

| Domain | 2021 | | 2022 | |
| --- | --- | --- | --- | --- |
| | Fractional Journals (Percent) | Fractional Papers (Percent) | Fractional Journals (Percent) | Fractional Papers (Percent) |
| **Life Sciences** | 129.67 (36.85%) | 533.50 (22.04%) | 138.50 (38.90%) | 618.67 (34.29%) |
| **Physical Sciences** | 81.00 (27.70%) | 626.50 (39.84%) | 90.33 (25.37%) | 575.17 (31.88%) |
| **Engineering & Technology** | 46.83 (15.49%) | 494.17 (32.51%) | 50.67 (14.23%) | 473.33 (26.24%) |
| **Social Sciences** | 40.33 (11.50%) | 71.00 (2.93%) | 49.33 (13.86%) | 85.00 (4.71%) |
| **Clinical, Pre-Clinical & Health** | 25.67 (8.22%) | 45.33 (2.65%) | 26.677 (7.49%) | 51.33 (2.85%) |
| **Arts & Humanities** | 0.5 (0.23%) | 0.5 (0.03%) | 0.5 (0.14%) | 0.50 (0.03%) |

## 5. Discussion

### 5.1 The evolution of citation formats in the R-language ecosystem

Our results offer, for the first time, a scanning and longitudinal analysis of citation formats in the R programming language ecosystem, one of the most influential open-source programming languages used by researchers (Li et al., 2017). We specifically used two longitudinal datasets, collected in 2021 and 2022, respectively, to analyze how R package citation formats change over time, as a first step towards a comprehensive understanding of the dynamic nature of software citation formats and the challenges to integrating software into the research system.

Our results show the existence of two parallel approaches to citing R packages: namely, using the software repository or a software publication. We find that only about 12% of all R packages have software papers as their citation format. However, there is an increasing, albeit slow, trend toward publishing R packages as software papers (and other types of academic outputs) and adopting the software paper as the official citation format. This shows that although not every R package with research value is published as a software paper, over time, more software developers are embracing the emerging genre of software paper and using it to promote their software in academic communities. More profoundly, this finding underscores a challenge in how research software should be traced in our research infrastructure. One notable example is that Clarivate's Data Citation Index only uses software repositories to trace research software (Park & Wolfram, 2019), leaving it unable to cover all published open-source software packages based on our findings.

Moreover, there are various irregularities in the software citation formats we examined, which makes it difficult to cite software and trace it within our existing research infrastructure. The most obvious issue is the frequent changes in software citation format even within the 15-month window selected for our research. Our results show that about 60% of *Remained* packages (those shown in both subsamples) have different citation formats in the two time points, among which 403 and 382 packages even have a different cited object or cited title, respectively. Software version, another important identifying metadata element for software citation, also significantly contributes to the changes in citation formats.

Building upon our previous study (Li, Chen, & Yan, 2019), our research demonstrates the extent to which citation format changes occur in a major open-source programming language ecosystem. This issue makes it very difficult, if not impossible, to use automatic methods to trace software cited in standard research databases, where it is assumed that (1) citations are static and (2) citations and cited objects only have one-to-one relationships. Even though not every changed metadata element in the citation format is equally important for the identification of software, or even the composition of the citation format, such changes pose a serious issue to be solved in the future so that an infrastructure to support software citation (and to a lesser degree, data citation) can be established.

Another frequently changed metadata field is the publication year of software. Our results show that after a software package is published into an academic output, the publication year is largely fixed (though changes can still happen). However, the publication year of a software repository can easily change when a new version of software is introduced. In quantitative science studies, it is a frequent practice to use the publication year of cited objects for various purposes, one of which is to calculate or estimate novelty (Burton & Kebler, 1960; Van Raan, 2004; Leydesdorff, 2009). However, our results demonstrate that such practices may not be effective for software entities, which have flexible publication years. Given the centrality of research methods to scientific innovation, this is an important issue to consider in future studies on innovation.

**5.2 Research disciplines of R software papers**

In this research, we also examined software papers published for R packages, as a first step to understand the practice of publishing software papers that has gradually emerged during the past few years. Although the share of R packages with a published software paper is still relatively low within the overall R-language community, we identified over 1,000 such papers, and the number increased by more than 100 from 2021 to 2022 (including the formal publication of preprints).

We specifically examined the journals and disciplines of R software papers as indexed in the Web of Science. Our results show that R software papers are published in both journals dedicated to software papers and those that accept software papers along with research articles. This is also the situation reported for the new academic genre of data papers (Candela et al., 2015), which makes it more difficult to trace software publishing activities through the bibliographic universe. Moreover, some of the top journals to publish R papers are those highly specialized journals that are not indexed in major research databases, most notably *The R Journal* and *R News*. This shows that some software development activities are only recorded in venues that cannot be observed using standard databases.

We further mapped the journals to the six GIPP research domains using the Web of Science classification system. Results show that even though the majority of software papers fall into STEM domains, computer science and engineering are not the sole center of these publications; more applied fields, such as life and physical sciences, also heavily contribute to the publication of software papers. This finding partially echoes prior results showing that software papers are published in a broad range of knowledge domains (Li, Chen, & Fang, 2019). From the perspective of sociology of science, software functions as a critical boundary object between actors in scientific collaboration (Hocquet & Wieber, 2018). We hope that our research is the first step towards a better understanding of the complex relationship between research software and scientific collaboration.

**6. Conclusion**

In this study, we analyzed the composition and evolution of citation formats of R-language packages deposited in the CRAN repository, the official and largest software package repository for the R programming language. We collected a longitudinal dataset of the official citation format of all R packages in May 2021 and August 2022 and compared how the packages and their citation formats changed over time. We specifically focused on the cited objects and the metadata fields in the citation format. Moreover, we examined the journals and research disciplines of software papers being cited in the citation format.

We find that software repository pages and software papers are the two most popular way to cite software. This is a confirmation, using a large-scale dataset, of the complexities of software citation from the perspective of citation format. Beyond the diversity of citation format, we also find that the citation format of the same software entity can easily change, even on the 15-month timescale that is examined in this research. Together, these two findings suggest that we will need a distinct method and research infrastructure to trace how software is cited in research outputs, one which considers the existence of and shifts among multiple citation formats that may exist for the same software object.

We also find that the size and share of R packages using a software paper have been increasing over time. This confirms the rise in the practice of publishing software papers that can lead to more recognition for the software developers and the construction of a more transparent research system. Moreover, for the first time, our research evaluates the disciplines within which software papers are published. Our results demonstrate that the majority of R software papers are published in STEM fields, though not necessarily computer science specifically. This shows the potential heterogeneity of software development activities in academia and could serve as an important baseline for future studies on the disciplinarity of software development and reuse activities.

Despite our timely contributions, our results have some limitations that should be considered by the readers. First, we only collected our data from the R ecosystem across two years; it represents only a small slice of the history of this single open-source programming language project. Despite the importance of R in research and data science (Loukides, 2010), other programming languages and software, especially Python, also exert strong impacts on research activities. We plan to expand our efforts to other research software in our next step, to fully understand how software entities are contributing to and shaping research. Additionally, this paper presents only a preliminary scanning of software papers. We also plan to analyze the publication of software papers more systematically, using findings from this analysis.

**References**

Barker, M., Chue Hong, N. P., Katz, D. S., Lamprecht, A.-L., Martinez-Ortiz, C., Psomopoulos, F., Harrow, J., Castro, L. J., Gruenpeter, M., Martinez, P. A., & Honeyman, T. (2022). Introducing the FAIR Principles for research software. *Scientific Data*, *9*(1), Article 1. https://doi.org/10.1038/s41597-022-01710-x

Borgman, C. L., Wallis, J. C., & Mayernik, M. S. (2012). Who's Got the Data? Interdependencies in science and technology collaborations. *Computer Supported Cooperative Work (CSCW)*, *21*(6), 485–523. https://doi.org/10.1007/s10606-012-9169-z

Bouquin, D. R., Chivvis, D. A., Henneken, E., Lockhart, K., Muench, A., & Koch, J. (2020). Credit lost: two decades of software citation in astronomy. *The Astrophysical Journal Supplement Series*, *249*(1), 8. https://doi.org/10.3847/1538-4365/ab7be6

Branstetter, L. G., Glennon, B., & Jensen, J. B. (2019). The IT revolution and the globalization of R&D. *Innovation Policy and the Economy*, *19*, 1–37. https://doi.org/10.1086/699931

Burton, R. E., & Kebler, R. W. (1960). The "half-life" of some scientific and technical literatures. *American Documentation*, *11*(1), 18–22. https://doi.org/10.1002/asi.5090110105

Candela, L., Castelli, D., Manghi, P., & Tani, A. (2015). Data journals: A survey. *Journal of the Association for Information Science and Technology*, *66*(9), 1747–1762. https://doi.org/10.1002/asi.23358

Chassanoff, A., & Altman, M. (2020). Curation as "Interoperability with the Future": Preserving scholarly research software in academic libraries. *Journal of the Association for Information Science & Technology*, *71*(3), 325–337. https://doi.org/10.5703/1288284315651

Chue Hong, N., Hole, B., & Moore, S. (2013). Software Papers: Improving the reusability and sustainability of scientific software. *Figshare. Journal Contribution*. https://doi.org/10.6084/M9.FIGSHARE.795303.V1

Druskat, S. (2020). Software and dependencies in research citation graphs. *Computing in Science & Engineering*, *22*(2), 8–21. https://doi.org/10.1109/MCSE.2019.2952840

Du, C., Cohoon, J., Lopez, P., & Howison, J. (2021). Softcite dataset: A dataset of software mentions in biomedical and economic research publications. *Journal of the Association for Information Science and Technology*, *72*(7), 870–884. https://doi.org/10.1002/asi.244 54

Du, C., Cohoon, J., Lopez, P., & Howison, J. (2022). Understanding progress in software citation: A study of software citation in the CORD-19 corpus. *PeerJ Computer Science*, *8*, e1022. https://doi.org/10.7717/peerj-cs.1022

Edwards, P. N., Jackson, S. J., Chalmers, M. K., Bowker, G. C., Borgman, C. L., Ribes, D., Burton, M., & Calvert, S. (2013). *Knowledge Infrastructures: Intellectual Frameworks and Research Challenges*. https://escholarship.org/uc/item/2mt6j2mh

Fox, J., & Leanage, A. (2016). R and the Journal of Statistical Software. *Journal of Statistical Software*, *73*, 1–13. https://doi.org/10.18637/jss.v073.i02

Gentleman, R. C., Carey, V. J., Bates, D. M., Bolstad, B., Dettling, M., Dudoit, S., Ellis, B., Gautier, L., Ge, Y., Gentry, J., & others. (2004). Bioconductor: Open software development for computational biology and bioinformatics. *Genome Biology*, *5*(10), 1–16.

Hocquet, A., & Wieber, F. (2018). Mailing list archives as useful primary sources for historians: Looking for flame wars. *Internet Histories*, 2(1-2), 38-54. https://www.tandfonline.com/doi/abs/10.1080/24701475.2018.1456741

Hong, C., Allen, A., Gonzalez-Beltran, A., de Waard, A., Smith, A. M., Robinson, C., Jones, C., Bouquin, D., Katz, D. S., Kennedy, D., Ryder, G., Hausman, J., Hwang, L., Jones, M. B., Harrison, M., Crosas, M., Wu, M., Löwe, P., Haines, R., … Pollard, T. (2019a). *Software Citation Checklist for Authors*. Zenodo. https://doi.org/10.5281/zenodo.3479199

Hong, C., Allen, A., Gonzalez-Beltran, de Waard, A., Smith, A. M., Robinson, C., Jones, C., Bouquin, D., Katz, D. S., Kennedy, D., Ryder, G., Hausman, J., Hwang, L., Jones, M. B., Harrison, M., Crosas, M., Wu, M., Löwe, P., Haines, R., … Pollard, T. (2019b). *Software Citation Checklist for Developers*. Zenodo. https://doi.org/10.5281/zenodo.3482769

Hornik, K. (2012). The Comprehensive R Archive Network. *WIREs Computational Statistics*, *4*(4), 394–398. https://doi.org/10.1002/wics.1212

Howison, J., & Bullard, J. (2015). Software in the scientific literature: Problems with seeing, finding, and using software mentioned in the biology literature. *Journal of the Association for Information Science and Technology*, *67*(9), 2137–2155. https://doi.org/10.1002/asi.23538

Howison, J., Deelman, E., McLennan, M. J., Ferreira da Silva, R., & Herbsleb, J. D. (2015). Understanding the scientific software ecosystem and its impact: Current and future measures. *Research Evaluation*, *24*(4), 454–470. https://doi.org/10.1093/reseval/rvv014

Istrate, A.-M., Li, D., Taraborelli, D., Torkar, M., Veytsman, B., & Williams, I. (2022). *A large dataset of software mentions in the biomedical literature* (arXiv:2209.00693). arXiv. https://doi.org/10.48550/arXiv.2209.00693

Jay, C., Haines, R., & Katz, D. S. (2021). Software must be recognised as an important output of scholarly research. *International Journal of Digital Curation*, *16*(1), 6. https://doi.org/10.2218/ijdc.v16i1.745

Katz, D. S., Hong, N. P. C., Clark, T., Muench, A., Stall, S., Bouquin, D., Cannon, M., Edmunds, S., Faez, T., Feeney, P., Fenner, M., Friedman, M., Grenier, G., Harrison, M., Heber, J., Leary, A., MacCallum, C., Murray, H., Pastrana, E., … Yeston, J. (2021). Recognizing the value of software: A software citation guide. *F1000Research*, *9:1257*. https://f1000research.com/articles/9-1257

Katz, D. S., Niemeyer, K. E., Smith, A. M., Anderson, W. L., Boettiger, C., Hinsen, K., Hooft, R., Hucka, M., Lee, A., Löffler, F., Pollard, T., & Rios, F. (2016). Software vs. Data in the context of citation. *PeerJ Preprints 4*: e2630v1. https://doi.org/10.7287/peerj.Preprints.2630v1

Kelley, A., & Garijo, D. (2021). A framework for creating knowledge graphs of scientific software metadata. *Quantitative Science Studies*, *2*(4), 1423–1446. https://doi.org/10.1162/qss_a_00167

Kelty, C. M. (2001, December 3). *Free software/free science*. First Monday. https://firstmonday.org/ojs/index.php/fm/article/download/902/811?inline=1

Kratz, J., & Strasser, C. (2014). Data publication consensus and controversies. *F1000Research*, 3:94. http://dx.doi.org/10.12688/f1000research.3979.3

Lamprecht, A.L., Garcia, L., Kuzak, M., Martinez, C., Arcila, R., Martin Del Pico, E., Dominguez Del Angel, V., van de Sandt, S., Ison, J., Martinez, P. A., McQuilton, P., Valencia, A., Harrow, J., Psomopoulos, F., Gelpi, J. L., Chue Hong, N., Goble, C., & Capella-Gutierrez, S. (2020). Towards FAIR principles for research software. *Data Science*, *3*(1), 37–59. https://doi.org/10.3233/DS-190026

LaZerte, S. (2021). *How to Cite R and R Packages*. https://ropensci.org/blog/2021/11/16/how-to-cite-r-and-r-packages/

Leydesdorff, L. (2009). How are new citation-based journal indicators adding to the bibliometric toolbox? *Journal of the American Society for Information Science and Technology*, *60*(7), 1327–1336. https://doi.org/10.1002/asi.21024

Li, K., Chen, P.Y., & Fang, Z. (2019). Disciplinarity of software papers: A preliminary analysis. *Proceedings of the Association for Information Science and Technology*, *56*(1), 706–708. https://doi.org/10.1002/pra2.143

Li, K., Chen, P.Y., & Yan, E. (2019). Challenges of measuring software impact through citations: An examination of the lme4 R package. *Journal of Informetrics*, *13*(1), 449–461.

Li, K., & Yan, E. (2018). Co-mention network of R packages: Scientific impact and clustering structure. *Journal of Informetrics*, *12*(1), 87–100.

Li, K., Yan, E., & Feng, Y. (2017). How is R cited in research outputs? Structure, impacts, and citation standard. *Journal of Informetrics*, *11*(4), 989–1002.

Loukides, M. (2010, June 2). *What is data science?* O'Reilly Media. https://www.oreilly.com/radar/what-is-data-science/

Manghi, P., Mannocci, A., Osborne, F., Sacharidis, D., Salatino, A., & Vergoulis, T. (2021). New trends in scientific knowledge graphs and research impact assessment. *Quantitative Science Studies*, *2*(4), 1296–1300. https://doi.org/10.1162/qss_e_00160

Manovich, L. (2013). *Software Takes Command*. Bloomsbury Academic. https://www.academia.edu/542750/Software_Takes_Command

Pan, X., Yan, E., Cui, M., & Hua, W. (2018). Examining the usage, citation, and diffusion patterns of bibliometric mapping software: A comparative study of three tools. *Journal of Informetrics*, *12*(2), 481–493.

Pan, X., Yan, E., & Hua, W. (2016). Disciplinary differences of software use and impact in scientific literature. *Scientometrics*, *109*(3), 1593–1610.

Park, H., & Wolfram, D. (2019). Research software citation in the Data Citation Index: Current practices and implications for research software sharing and reuse. *Journal of Informetrics*, *13*(2), 574–582. https://doi.org/10.1016/j.joi.2019.03.005

Parsons, M., & Fox, P. (2013). Is Data Publication the Right Metaphor? *Data Science Journal*, *12*(0), Article 0. https://doi.org/10.2481/dsj.WDS-042

Schindler, D., Bensmann, F., Dietze, S., & Krüger, F. (2022). The role of software in science: A knowledge graph-based analysis of software mentions in PubMed Central. *PeerJ Computer Science*, *8*, e835. https://doi.org/10.7717/peerj-cs.835

Schindler, D., Zapilko, B., & Krüger, F. (2020). Investigating software usage in the social sciences: A knowledge graph approach. *In: European Semantic Web Conference. ESWC 2020. Lecture Notes in Computer Science*, 271–286, Heraklion (Crete, Greece), Springer, Cham. https://doi.org/10.1007/978-3-030-49461-2_16

Shu, F., Julien, C.A., Zhang, L., Qiu, J., Zhang, J., & Larivière, V. (2019). Comparing journal and paper level classifications of science. *Journal of Informetrics*, *13*(1), 202–225. https://doi.org/10.1016/j.joi.2018.12.005

Smith, A. M., Katz, D. S., & Niemeyer, K. E. (2016). Software citation principles. *PeerJ Computer Science*, *2*, e86. https://doi.org/10.7717/peerj-cs.86

United Nations Conference on Trade and Development. (2012). Software for development. In United Nations Conference on Trade and Development, *Information Economy Report 2012* (pp. 1–16). UN. https://doi.org/10.18356/56e6e4ed-en

Van Raan, A. F. J. (2004). Sleeping Beauties in science. *Scientometrics*, *59*(3), 467–472. https://doi.org/10.1023/B:SCIE.0000018543.82441.f1

Wang, Y., & Zhang, C. (2020). Using the full-text content of academic articles to identify and evaluate algorithm entities in the domain of natural language processing. *Journal of Informetrics*, *14*(4), 101091. https://doi.org/10.1016/j.joi.2020.101091

Wilkinson, M. D., Dumontier, M., Aalbersberg, Ij. J., Appleton, G., Axton, M., Baak, A., Blomberg, N., Boiten, J.-W., da Silva Santos, L. B., Bourne, P. E., Bouwman, J., Brookes, A. J., Clark,

T., Crosas, M., Dillo, I., Dumon, O., Edmunds, S., Evelo, C. T., Finkers, R., … Mons, B. (2016). The FAIR Guiding Principles for scientific data management and stewardship. *Scientific Data*, *3*(1), Article 1. https://doi.org/10.1038/sdata.2016.18

Wolfram, S. (1984). Computer Software in Science and Mathematics. *Scientific American*, *251*(3), 188–203.